\documentclass[aps,twocolumn,showpacs,preprintnumbers,amsmath,amssymb,
nofootinbib,showkeys,floatfix]{revtex4-1}

\usepackage{epsfig}
\usepackage{graphicx}
\usepackage{dcolumn}
\usepackage{latexsym}
\usepackage{graphicx}
\usepackage{enumerate}
\usepackage{float}
\usepackage{placeins}
\usepackage{supertabular}

\usepackage{hyperref} 
\hypersetup{
   colorlinks=true,
   linkcolor=blue,
   citecolor=blue,
   urlcolor=blue
}

\def\tstrut{\vrule height2.25ex depth0pt width0pt} 

\bibpunct{[}{]}{,}{n}{,}{,} 

\begin{document}

\title{Nonleptonic $B\to D^{(\ast)}D_{sJ}^{(\ast)}$ decays and the nature of
the orbitally excited charmed-strange mesons}
\author{J. Segovia}
\author{C. Albertus}
\author{E. Hern\'andez}
\author{F. Fern\'andez}
\author{D.R. Entem}
\affiliation{Grupo de F\'isica Nuclear e IUFFyM, \\ Universidad de 
Salamanca, E-37008 Salamanca, Spain}
\date{\today}

\begin{abstract}
The Belle Collaboration has recently reported a study of the decays $B \to
D_{s1}(2536)^{+}\bar{D}^{(\ast)}$ and has given also estimates of relevant
ratios between branching fractions of decays $B \to
D^{(\ast)}D_{sJ}^{(\ast)}$ providing important information to check the
structure of the $D_{s0}^{\ast}(2317)$, $D_{s1}(2460)$ and $D_{s1}(2536)$
mesons. The disagreement between experimental data and Heavy Quark Symmetry has
been used as an indication that $D_{s0}^{\ast}(2317)$ and $D_{s1}(2460)$ mesons
could have a more complex structure than the canonical $c\bar{s}$ one. We
analyze these ratios within the framework of a constituent quark model, which
allows us to incorporate the effects given by finite $c$-quark mass corrections.
Our findings are that while the $D_{s1}(2460)$ meson could have a sizable
non-$q\bar{q}$ component, the $D_{s0}^{\ast}(2317)$ and $D_{s1}(2536)$ mesons
seem to be well described by a pure $q\bar{q}$ structure.
\end{abstract}


\maketitle

\section{INTRODUCTION}
\label{sec:introduction}

$B$-factories, through $e^{+}e^{-}$ collisions, tuned to different bottomonium
energy resonances, have become a source of data on heavy hadrons. Bottomonium
states decay most of the cases to $B\bar{B}$ pairs, and these $B$ mesons decay
subsequently into charmed and charmless hadrons via the weak interaction.

Nonleptonic decays of $B$ mesons, described at quark level by an effective
four-quark interaction $\bar{b} \to \bar{c}c\bar{s}$, have been used to search
for new charmonium and charmed-strange mesons and to study their properties in
detail. Within the charmed-strange sector, the BaBar Collaboration found,  in
the inclusive $D_{s}^{+}\pi^{0}$ invariant mass distribution from $e^{+}e^{-}$
annihilation data, the narrow state
$D_{s0}^{\ast}(2317)$~\cite{PhysRevLett.90.242001}. The CLEO Collaboration,
aiming to confirm the previous state, observed its doublet partner
$D_{s1}(2460)$ in the $D_{s}^{\ast+}\pi^{0}$ final
state~\cite{PhysRevD.68.032002}. However, the properties of these states were
not well known until the Belle Collaboration observed the $B \to
\bar{D}D_{s0}^{\ast}(2317)$ and $B \to \bar{D}D_{s1}(2460)$
decays~\cite{PhysRevLett.91.262002}.

First observations of the $B \to \bar{D}^{(\ast)}D_{s1}(2536)$ decay modes have
been reported by the BaBar
Collaboration~\cite{PhysRevD.74.091101,PhysRevD.77.011102} and an upper
limit on the decay $B^{0} \to D^{\ast-}D_{s1}(2536)^{+}$ was also obtained by
the Belle Collaboration~\cite{PhysRevD.76.072004}. The most recent analysis of
the production of $D_{s1}(2536)^{+}$ in double charmed $B$ meson decays has been
reported by the Belle Collaboration in Ref.~\cite{PhysRevD.83.051102}. Using the
latest measurements of the $B \to D^{(\ast)}D_{sJ}^{(\ast)}$ branching
fractions~\cite{PDG2010} they calculate the ratios
\begin{equation*}
\begin{split}
R_{D0} &= \frac{{\cal B}(B \to DD_{s0}^{\ast}(2317))}{{\cal B}(B \to
DD_{s})} = 0.10\pm0.03, \\
R_{D^{\ast}0} &= \frac{{\cal B}(B \to D^{\ast}D_{s0}^{\ast}(2317))}{{\cal B}(B
\to D^{\ast}D_{s})} = 0.15\pm0.06,
\end{split}
\end{equation*}

\begin{equation}
\begin{split}
R_{D1} &= \frac{{\cal B}(B \to DD_{s1}(2460))}{{\cal B}(B \to
DD_{s}^{\ast})} = 0.44\pm0.11, \\
R_{D^{\ast}1} &= \frac{{\cal B}(B \to D^{\ast}D_{s1}(2460))}{{\cal B}(B \to
D^{\ast}D_{s}^{\ast})} = 0.58\pm0.12.
\label{eq:DDsratio1}
\end{split}
\end{equation}
In addition, the same ratios are calculated for $B \to
D^{(\ast)}D_{s1}(2536)^{+}$ decays, using combined results by the
BaBar~\cite{PhysRevD.77.011102} and Belle~\cite{PhysRevD.83.051102} 
Collaborations
\begin{equation}
\begin{split}
R_{D1'} &= \frac{{\cal B}(B \to DD_{s1}(2536))}{{\cal B}(B \to
DD_{s}^{\ast})} = 0.049\pm0.010, \\
R_{D^{\ast}1'} &= \frac{{\cal B}(B \to D^{\ast}D_{s1}(2536))}{{\cal B}(B \to
D^{\ast}D_{s}^{\ast})} = 0.044\pm0.010.
\label{eq:DDsratio2}
\end{split}
\end{equation}

>From a theoretical point of view, this kind of decays can be described using the
factorization approximation~\cite{PhysRevD.73.054024}. This amounts to evaluate
the matrix element which describes the $B \to D^{(\ast)}D_{sJ}^{(\ast)}$ weak
decay process as a product of two matrix elements, one for the $B$ weak
transition into the $D^{(\ast)}$ meson and the second one for the weak
creation of the $c\bar{s}$ pair which makes the $D_{sJ}^{(\ast)}$ meson. The
latter matrix element is proportional to the corresponding $D_{sJ}^{(\ast)}$
meson decay constant and therefore these processes can give information about
the $D_{sJ}$ structure.

The $D_{s0}^{\ast}(2317)$ and $D_{s1}(2460)$ states belong to the doublet
$j_{q}^{P}=\frac{1}{2}^{+}$ predicted by Heavy Quark Symmetry (HQS). They have
surprisingly light masses and are located below $DK$ and $D^{\ast}K$ thresholds,
respectively. This implies that these states are narrow. Neither quark models
nor Lattice QCD calculations~\cite{mohler2011d} can give a satisfactory
explanation of this doublet in the charmed and charmed-strange sectors. This
fact has stimulated a fruitful line of research, suggesting that their structure
is much richer than what one might guess assuming the $q\bar{q}$ picture, which
has proved quite successful in other sectors. Some authors have suggested a $qq
\bar{q}\bar{q}$ structure~\cite{PhysRevD.68.054006,PhysRevLett.91.012003} or
mixing between the usual $q\bar{q}$ structure and
four quark~\cite{PhysRevD.73.034002}. Also there have been suggestions that
these states might be molecular states. However, one must mention that the
${\cal B}(B\to D_{s0}^{\ast}(2317)K)$ is the same order of magnitude as ${\cal
B}(B\to D_{s}K)$ and at least a factor of $2$ larger than the ${\cal B}(B\to
D_{s1}(2460)K)$ branching fraction~\cite{PhysRevLett.94.061802}, in contrast
with the naive expectation that decays within the same spin doublet would have
similar rates. This fact suggests that the $D_{s0}^{\ast}(2317)$ meson may be a
conventional $0^{+}$ $c\bar{s}$ state and the $D_{s1}(2460)$ has a more complex
structure.

In any case the study of nonleptonic decays can help to shed light on the
structure of the P-wave open charm mesons

The $D_{sJ}^{(\ast)}$ meson decay constants are not known experimentally
except for the ground state, $D_{s}$, which has been measured by different
collaborations. Another way to study $D_{sJ}^{(\ast)}$ mesons that does not rely
on the knowledge of their decay constants is through the decays $B_{s}\to
D_{sJ}^{(\ast)}M$ where $M$ is a meson with a well known decay constant.
However, the experimental study of these processes is currently difficult
leaving, for the time being, the $B \to D^{(\ast)}D_{sJ}^{(\ast)}$ decay
processes as our best option to study $D_{sJ}^{(\ast)}$ meson properties.

According to Ref.~\cite{Datta2003164}, within the factorization approximation
and in the heavy quark limit, the ratios $R_{D0}$ and $R_{D1}$ can be written
as
\begin{equation}
\begin{split}
R_{D0} &= R_{D^{\ast}0} = \left|
\frac{f_{D_{s0}^{\ast}(2317)}}{f_{D_{s}}}\right|^{2}, \\
R_{D1} &= R_{D^{\ast}1} = \left|
\frac{f_{D_{s1}(2460)}}{f_{D_{s}^{\ast}}}\right|^{2},
\end{split}
\end{equation}
where the phase space effects are neglected because they are subleading in the
heavy quark expansion. Now, in the heavy quark limit one has
$f_{D_{s0}^{\ast}}=f_{D_{s1}}$ and $f_{D_{s}} = f_{D_{s}^{\ast}}$ and so one
would predict $R_{D0} \approx R_{D1}$. Moreover, there are several estimates of
the decay constants, always in the heavy quark
limit~\cite{Yaouanc200159,Colangelo2002193,PhysRevD.60.033002}, that predict for
$P$-wave $j_{q}=1/2$ states similar decay constants as for the ground state
mesons (i.e. $D_{s}$ and $D_{s}^{\ast}$), and very small decay constants for
$P$-wave, $j_{q}=3/2$ states. Then, these approximations lead to ratios of order
one for $D_{s0}^{\ast}(2317)$ and $D_{s1}(2460)$, in strong disagreement with
experiment.

Leaving aside that the factorization approximation has been recently analyzed in
Refs.~\cite{PhysRevD.64.094001,PhysRevD.58.014007,PhysRevD.55.6780} finding that
it works well in these kind of processes, we will concentrate in the influence
of the effect of the finite $c$-quark mass in the theoretical predictions. As
found in Ref.~\cite{PhysRevD.72.094010}, $1/m_{Q}$ contributions could give
large corrections to various quantities describing $B\to D^{\ast\ast}$
transitions and we expect they could also play an important role in this case.

We work within the framework of the constituent quark model described in
Ref.~\cite{vijande2005constituent} and which is widely used in hadronic
spectroscopy~\cite{vijande2005constituent,PhysRevC.64.058201}. It has recently
been applied to mesons containing heavy quarks in
Refs.~\cite{PhysRevD.78.114033,PhysRevD.80.054017}. It has also been used
successfully in the description of semileptonic $B \to D^{\ast\ast}$ and $B_{s}
\to D_{s}^{\ast\ast}$ decays in Ref.~\cite{segovia2011semileptonic}, which
provides us with confidence that the model describes well the weak decay
transitions $B \to D^{(\ast)}$ which is one of the terms in the calculation of 
$B \to D^{(\ast)}D_{sJ}^{(\ast)}$ decays.

The paper is organized as follows: In Sec.~\ref{sec:CQM}, we introduce the
constituent quark model and discuss its predictions in the charmed and
charmed-strange sectors. Sec.~\ref{sec:weakdecays} is devoted to explain the
theoretical framework through which we calculate the $B \to
D^{(\ast)}D_{sJ}^{(\ast)}$ decays. Finally, we present our results in
Sec.~\ref{sec:results} and give the conclusions in Sec.~\ref{sec:conclusions}.

\section{CONSTITUENT QUARK MODEL}
\label{sec:CQM}

Constituent light quark masses and Goldstone-boson exchanges coming from the
spontaneous chiral symmetry breaking of the QCD Lagrangian, together with the
perturbative one-gluon exchange (OGE) and the nonperturbative confining
interaction are the main pieces of potential models. Using this idea, Vijande
{\it et al.}~\cite{vijande2005constituent} developed a model of the quark-quark
interaction which is able to describe meson phenomenology from the light to the
heavy quark sector.

A consistent description of light, strange and heavy mesons requires an
effective scale-dependent strong coupling constant. We use the frozen coupling
constant of Ref.~\cite{vijande2005constituent}
\begin{equation}
\alpha_{s}(\mu_{ij})=\frac{\alpha_{0}}{\ln\left(
\frac{\mu_{ij}^{2}+\mu_{0}^{2}}{\Lambda_{0 }^{2}} \right)},
\end{equation}
where $\mu_{ij}$ is the reduced mass of the $q\bar{q}$ pair, and $\alpha_{0}$,
$\mu_{0}$ and $\Lambda_{0}$ are parameters of the model determined by a global 
fit to all meson spectrum.

In the heavy quark sector chiral symmetry is explicitly broken and
Goldstone-boson exchanges do not appear. Thus, OGE and confinement are the only
interactions remaining. 

The one-gluon exchange potential is generated from the vertex Lagrangian
\begin{equation}
{\mathcal L}_{qqg} = i\sqrt{4\pi\alpha_{s}} \bar{\psi} \gamma_{\mu} G^{\mu}_{c}
\lambda^{c} \psi,
\label{Lqqg}
\end{equation}
where $\lambda^{c}$ are the $SU(3)$ color matrices and $G^{\mu}_{c}$ is the
gluon field. The resulting potential contains central, tensor and spin-orbit
contributions given by
\begin{widetext}
\begin{equation}
\begin{split}
&
V_{\rm OGE}^{\rm C}(\vec{r}_{ij}) =
\frac{1}{4}\alpha_{s}(\vec{\lambda}_{i}^{c}\cdot
\vec{\lambda}_{j}^{c})\left[ \frac{1}{r_{ij}}-\frac{1}{6m_{i}m_{j}} 
(\vec{\sigma}_{i}\cdot\vec{\sigma}_{j}) 
\frac{e^{-r_{ij}/r_{0}(\mu)}}{r_{ij}r_{0}^{2}(\mu)}\right], \\
& 
V_{\rm OGE}^{\rm T}(\vec{r}_{ij})=-\frac{1}{16}\frac{\alpha_{s}}{m_{i}m_{j}}
(\vec{\lambda}_{i}^{c}\cdot\vec{\lambda}_{j}^{c})\left[ 
\frac{1}{r_{ij}^{3}}-\frac{e^{-r_{ij}/r_{g}(\mu)}}{r_{ij}}\left( 
\frac{1}{r_{ij}^{2}}+\frac{1}{3r_{g}^{2}(\mu)}+\frac{1}{r_{ij}r_{g}(\mu)}\right)
\right]S_{ij}, \\
&
\begin{split}
V_{\rm OGE}^{\rm SO}(\vec{r}_{ij})= &  
-\frac{1}{16}\frac{\alpha_{s}}{m_{i}^{2}m_{j}^{2}}(\vec{\lambda}_{i}^{c} \cdot
\vec{\lambda}_{j}^{c})\left[\frac{1}{r_{ij}^{3}}-\frac{e^{-r_{ij}/r_{g}(\mu)}}
{r_{ij}^{3}} \left(1+\frac{r_{ij}}{r_{g}(\mu)}\right)\right] \times \\ & \times 
\left[((m_{i}+m_{j})^{2}+2m_{i}m_{j})(\vec{S}_{+}\cdot\vec{L})+
(m_{j}^{2}-m_{i}^{2}) (\vec{S}_{-}\cdot\vec{L}) \right],
\end{split}
\end{split}
\end{equation}
\end{widetext}
where $r_{0}(\mu_{ij})=\hat{r}_{0}\frac{\mu_{nn}}{\mu_{ij}}$ and
$r_{g}(\mu_{ij})=\hat{r}_{g}\frac{\mu_{nn}}{\mu_{ij}}$ are regulators. Note the
contact term of the central part of the one-gluon exchange potential has been
regularized as follows
\begin{equation}
\delta(\vec{r}_{ij})\sim\frac{1}{4\pi
r_{0}^{2}}\frac{e^{-r_{ij}/r_{0}}}{r_{ij}}.
\label{eq:delta}
\end{equation}

Although there is no analytical proof, it is a general belief that confinement
emerges from the force between the gluon color charges. When two quarks are
separated, due to the non-Abelian character of the theory, the gluon fields
self-interact forming color strings which bring the quarks together.

In a pure gluon gauge theory the potential energy of the $q\bar{q}$ pair grows 
linearly with the quark-antiquark distance. However, in full QCD the presence of
sea quarks may soften the linear potential, due to the screening of the color
charges, and eventually leads to the breaking of the string. We incorporate this
feature in our confinement potential as
\begin{equation}
\begin{split}
&
V_{\rm CON}^{\rm C}(\vec{r}_{ij})=\left[-a_{c}(1-e^{-\mu_{c}r_{ij}})+\Delta
\right] (\vec{\lambda}_{i}^{c}\cdot\vec{\lambda}_{j}^{c}), \\
&
\begin{split}
&
V_{\rm CON}^{\rm SO}(\vec{r}_{ij}) =
-(\vec{\lambda}_{i}^{c}\cdot\vec{\lambda}_{j}^{c}) \frac{a_{c}\mu_{c}e^{-\mu_{c}
r_{ij}}}{4m_{i}^{2}m_{j}^{2}r_{ij}} \times \\
&
\times \left[((m_{i}^{2}+m_{j}^{2})(1-2a_{s})+4m_{i}m_{j}(1-a_{s}))(\vec{S}_{+}
\cdot\vec{L}) \right. \\
&
\left. \quad\,\, +(m_{j}^{2}-m_{i}^{2}) (1-2a_{s}) (\vec{S}_{-}\cdot\vec{L})
\right],
\end{split}
\end{split}
\end{equation}
where $a_{s}$ controls the mixture between the scalar and vector Lorentz
structures of the confinement. At short distances this potential presents a
linear behavior with an effective confinement strength,
$\sigma=-a_{c}\,\mu_{c}\,(\vec{\lambda}^{c}_{i}\cdot \vec{\lambda}^{c}_{j})$,
while it becomes constant at large distances. This type of potential shows a
threshold defined by
\begin{equation}
V_{\rm thr}=\{-a_{c}+ \Delta\}(\vec{\lambda}^{c}_{i}\cdot
\vec{\lambda}^{c}_{j}).
\end{equation}
No $q\bar{q}$ bound states can be found for energies higher than this
threshold. The system suffers a transition from a color string configuration
between two static color sources into a pair of static mesons due to the
breaking of the color flux-tube and the most favored subsequent decay into
hadrons.

\begin{table}[!t]
\begin{center}
\begin{tabular}{ccc}
 \hline
 \hline
 \tstrut
 Quark masses    & $m_{n}$ (MeV) & $313$ \\
 		 & $m_{s}$ (MeV) & $555$ \\
 		 & $m_{c}$ (MeV) & $1763$ \\
 		 & $m_{b}$ (MeV) & $5110$ \\
\hline
 OGE & $\alpha_{0}$ & $2.118$ \\
     & $\Lambda_{0}$ $(\mbox{fm}^{-1})$ & $0.113$ \\
     & $\mu_{0}$ (MeV) & $36.976$ \\
     & $\hat{r}_{0}$ (fm) & $0.181$ \\
     & $\hat{r}_{g}$ (fm) & $0.259$ \\
\hline
 Confinement & $a_{c}$ (MeV) & $507.4$ \\
	       & $\mu_{c}$ $(\mbox{fm}^{-1})$ & $0.576$ \\
	       & $\Delta$ (MeV) & $184.432$ \\
	       & $a_{s}$ & $0.81$ \\
 \hline
 \hline
\end{tabular}
\caption{\label{tab:parameters} Quark model parameters.}
\end{center}
\end{table}

Among the different methods to solve the Schr\"odinger equation in order to 
find the quark-antiquark bound states, we use the Gaussian Expansion
Method~\cite{Hiyama:2003cu} which provides enough accuracy and it simplifies the
subsequent evaluation of the decay amplitude matrix elements.

This procedure provides the radial wave function solution of the Schr\"odinger
equation as an expansion in terms of basis functions
\begin{equation}
R_{\alpha}(r)=\sum_{n=1}^{n_{max}} c_{n}^\alpha \phi^G_{nl}(r),
\end{equation} 
where $\alpha$ refers to the channel quantum numbers. The coefficients,
$c_{n}^\alpha$, and the eigenvalue, $E$, are determined from the Rayleigh-Ritz
variational principle
\begin{equation}
\sum_{n=1}^{n_{max}} \left[\left(T_{n'n}^\alpha-EN_{n'n}^\alpha\right)
c_{n}^\alpha+\sum_{\alpha'}
\ V_{n'n}^{\alpha\alpha'}c_{n}^{\alpha'}=0\right],
\end{equation}
where $T_{n'n}^\alpha$, $N_{n'n}^\alpha$ and $V_{n'n}^{\alpha\alpha'}$ are the 
matrix elements of the kinetic energy, the normalization and the potential, 
respectively. $T_{n'n}^\alpha$ and $N_{n'n}^\alpha$ are diagonal, whereas the
mixing between different channels is given by $V_{n'n}^{\alpha\alpha'}$.

Following Ref.~\cite{Hiyama:2003cu}, we employ Gaussian trial functions with
ranges  in geometric progression. This enables the optimization of ranges
employing a small number of free parameters. Moreover, the geometric
progression is dense at short distances, so that it enables the description of
the dynamics mediated by short range potentials. The fast damping of the
gaussian tail does not represent an issue, since we can choose the maximal
range much longer than the hadronic size.

Model parameters fitted over the whole meson
spectra~\cite{vijande2005constituent,PhysRevD.78.114033} are shown in
Table~\ref{tab:parameters}.

The model described above is not able to reproduce the spectrum of the $P$-wave
charmed and charmed-strange mesons. The inconsistency with experiment is mainly
due to the fact that the mass splittings between the $D_{s0}^{\ast}(2317)$,
$D_{s1}(2460)$ and $D_{s1}(2536)$ mesons are not well reproduced. The same
problem appears in Lattice QCD calculations or other quark models
~\cite{mohler2011d}.

In order to improve these mass splittings we follow the proposal of
Ref.~\cite{Lakhina2007159} and include one-loop corrections to the OGE potential
as derived by Gupta {\it et al.}~\cite{PhysRevD.24.2309}. This corrections shows
a spin-dependent term which affects only mesons with different flavor quarks.

The net result is a quark-antiquark interaction that can be written as:
\begin{equation}
V(\vec{r}_{ij})=V_{\rm
OGE}(\vec{r}_{ij})+V_{\rm CON}(\vec{r}_{ij})+V_{\rm OGE}^{\rm 1-loop}
(\vec{r}_{ij}),
\end{equation}
where $V_{\rm OGE}$ and $V_{\rm CON}$ were defined before and are treated
non-perturbatively. $V_{\rm OGE}^{\rm 1-loop}$ is the one-loop correction to OGE
potential which is treated perturbatively. As in the case of $V_{\rm OGE}$ and
$V_{\rm CON}$, $V_{\rm OGE}^{\rm 1-loop}$ contains central, tensor and
spin-orbit contributions, given by~\cite{Lakhina2007159}
\begin{widetext}
\begin{equation}
\begin{split}
&
V_{OGE}^{\rm 1-loop,C}(\vec{r}_{ij})=0, \\
&
\begin{split}
V_{OGE}^{\rm 1-loop,T}(\vec{r}_{ij}) = \frac{C_{F}}{4\pi}
\frac{\alpha_{s}^{2}}{m_{i}m_{j}}\frac{1}{r^{3}}S_{ij} 
&
\left[\frac{b_{0}}{2}\left(\ln(\mu
r_{ij})+\gamma_{E}-\frac{4}{3}\right)+\frac{5}{12}b_ {0}-\frac{2}{3}C_{A}
\right. \\ & \left.
+\frac{1}{2}\left(C_{A}+2C_{F}-2C_{A}\left(\ln(\sqrt{m_{i}m_{j}}\,r_{ij}
)+\gamma_ {
E}-\frac{4}{3}\right)\right)\right],
\end{split} \\
&
\begin{split}
&V_{OGE}^{\rm 1-loop,SO}(\vec{r}_{ij})=\frac{C_{F}}{4\pi}
\frac{\alpha_{s}^{2}}{m_{i}^{2}m_{j}^{2}}\frac{1}{r^{3}}\times \\
&
\begin{split}
\times\Bigg\lbrace (\vec{S}_{+}\cdot\vec{L}) & \Big[
\left((m_{i}+m_{j})^{2}+2m_{i}m_{j}\right)\left(C_{F}+C_{A}-C_{A}
\left(\ln(\sqrt{m_{i}m_{j}}\,r_{ij})+\gamma_{E}\right)\right) \\
&
+4m_{i}m_{j}\left(\frac{b_{0}}{2}\left(\ln(\mu
r_{ij})+\gamma_{E}\right)-\frac{1}{12}b_{0}-\frac{1}{2}C_{F}-\frac{7}{6}C_{A}
+\frac{C_{A}}{2}\left(\ln(\sqrt{m_{i}m_{j}}\,r_{ij})+\gamma_{E}\right)\right)
\\
&
+\frac{1}{2}(m_{j}^{2}-m_{i}^{2})C_{A}\ln\left(\frac{m_{j}}{m_{i}}\right)\Big] 
\end{split} \\
&
\begin{split}
\,\,\,\,\,\,\,+(\vec{S}_{-}\cdot\vec{L}) &
\Big[(m_{j}^{2}-m_{i}^{2})\left(C_{F}+C_{A}-C_{A}\left(\ln(\sqrt{m_{i}m_{j}}\,r_
{ij})+\gamma_{E}\right)\right) \\
&
+\frac{1}{2}(m_{i}+m_{j})^{2}C_{A}\ln\left(\frac{m_{j}}{m_{i}}\right)\Big]
\Bigg\rbrace,
\end{split}
\end{split}
\end{split}
\end{equation}
\end{widetext}
where $C_{F}=4/3$, $C_{A}=3$, $b_{0}=9$, $\gamma_{E}=0.5772$ and the scale
$\mu\sim1\,{\rm GeV}$.

\begin{table*}[t!]
\begin{center}
\begin{tabular}{c|cccccccccccc}
\hline
\hline
\tstrut
& \multicolumn{12}{c}{Charmed} \\[1.5ex]
& \multicolumn{4}{c}{$\underline{j_q^P=1/2^-}$} & \multicolumn{4}{c}{$\underline{j_q^P=1/2^+}$} & \multicolumn{4}{c}{$\underline{j_q^P=3/2^+}$} \\[2ex]
& $0^{-}$ & & $1^{-}$ & & $0^{+}$ & & $1^{+}$ & & $1^{+}$ & & $2^{+}$ & \\
This work $(\alpha_{s})$ & $1896$ & & $2017$ & & $2516$ & & $2596$ & & $2466$ & & $2513$ & \\
This work $(\alpha_{s}^{2})$ & $1896$ & & $2014$ & & $2362$ & &$2535$  & & $2499$ & & $2544$ & \\
Exp. & $1867.7\pm0.3$ & & $2010.25\pm0.14$ & & $2403\pm38$ & & $2427\pm36$ & &  $2423.4\pm3.1$ & & $2460.1\pm4.4$ & \\
\hline
\hline
\tstrut
& \multicolumn{12}{c}{Charmed-strange} \\[1.5ex]
& \multicolumn{4}{c}{$\underline{j_q^P=1/2^-}$} & \multicolumn{4}{c}{$\underline{j_q^P=1/2^+}$} & \multicolumn{4}{c}{$\underline{j_q^P=3/2^+}$} \\[2ex]
& $0^{-}$ & & $1^{-}$ & & $0^{+}$ & & $1^{+}$ & & $1^{+}$ & & $2^{+}$ & \\
This work $(\alpha_{s})$ & $1984$ & & $2110$ & & $2510$ & & $2593$ & & $2554$ & & $2591$ & \\
This work $(\alpha_{s}^{2})$ & $1984$ & & $2104$ & & $2383$ & & $2570$ & & $2560$ & & $2609$ & \\
Exp. & $1969.0\pm1.4$ & & $2112.3\pm0.5$ & & $2318.0\pm1.0$  & & $2459.6\pm0.9$ & & $2535.12\pm0.25$ & & $2572.6\pm0.9$ & \\
\hline
\hline
\end{tabular}
\caption{\label{tab:1loopDmesons} Masses of well established charmed and
charmed-strange mesons predicted by the constituent quark model $(\alpha_{s})$
and those including one-loop corrections to the OGE potential
$(\alpha_{s}^{2})$.}
\end{center}
\end{table*}

Table~\ref{tab:1loopDmesons} shows the masses of well established charmed and
charmed-strange mesons predicted by the constituent quark model and those
including one-loop corrections to the OGE potential.

The masses predicted for the $0^{-}$ and $1^{-}$ states agree with the
experimental measurements in both sectors. These states can be identified with
the members of the lowest lying $j_{q}^{P}=\frac{1}{2}^{-}$ doublet predicted by
HQS.

The doublet $j_{q}^{P}=\frac{3}{2}^{+}$, which corresponds to the $2^{+}$ state
and one of the low lying $1^{+}$ states, is in reasonable agreement with
experiment. 

The charmed and charmed-strange $0^{+}$ states are sensitive to the one-loop
corrections of the OGE potential which bring their masses closer to experiment.
However, the spin dependent corrections to the OGE potential, are not enough to 
solve the puzzle in the $1^{+}$ sector. A possible explanation for the low mass
of this state has been proposed in Ref.~\cite{PhysRevD.80.054017} were the
$1^{+}$ $c\bar{s}$ states are coupled to a $c\bar sn\bar n$ tetraquark
structure. One finds that the $J^{P}=1^{+}$ $D_{s1}(2460)$ has an important
non-$q\bar{q}$ contribution whereas the $J^{P}=1^{+}$ $D_{s1}(2536)$ is almost a
pure $q\bar{q}$ state.

\section{NONLEPTONIC $B \to D^{(\ast)}D_{sJ}^{(\ast)}$ DECAYS}
\label{sec:weakdecays}

We give account of the nonleptonic decays $B \to D^{(\ast)}D_{sJ}^{(\ast)}$ with
$D^{(\ast)}$ the $D$ or $D^{\ast}$ mesons, and $D_{sJ}^{(\ast)}$ the mesons
$D_{s0}^{\ast}(2317)$, $D_{s1}(2460)$ and $D_{s1}(2536)$. These decay modes
involve a $b\to c$ transition at the quark level, governed by the effective
Hamiltonian~\cite{PhysRevD.74.074008,PhysRevD.73.054024,RevModPhys.68.1125}
\begin{equation}
H_{\rm eff.}=\frac{G_{F}}{\sqrt{2}}\,\left\lbrace V_{cb} \left[
C_{1}(\mu')\,Q_{1}^{cb}+C_{2}(\mu')\,Q_{2}^{cb} \right] + h.c.\right\rbrace,
\label{eq:Heff}
\end{equation}
in which penguin operators have been neglected. In Eq.~(\ref{eq:Heff}),
$C_{1}(\mu')$ and $C_{2}(\mu')$ are scale-dependent Wilson coefficients, being
$\mu' \simeq m_{b}$ the appropriate energy scale  in this case. The $Q_{1}^{cb}$
and $Q_{2}^{cb}$ are local four-quark operators given by
\begin{equation}
\begin{split}
Q_{1}^{cb} \! &= \! V_{cs}^{\ast}
\left[\bar{\psi}_{c}(0)\gamma_{\mu}({\cal I}-\gamma_{5})\psi_{b}(0)\right]\!\!
\left[\bar{\psi}_{s}(0)\gamma^{\mu}({\cal I}-\gamma_{5})\psi_{c}(0)\right], \\
Q_{2}^{cb} \! &= \! V_{cs}^{\ast}
\left[\bar{\psi}_{s}(0)\gamma_{\mu}({\cal I}-\gamma_{5})\psi_{b}(0)\right]\!\!
\left[\bar{\psi}_{c}(0)\gamma^{\mu}({\cal I}-\gamma_{5})\psi_{c}(0)\right].
\end{split}
\end{equation}

\begin{figure*}[!t]
\begin{center}
\epsfig{figure=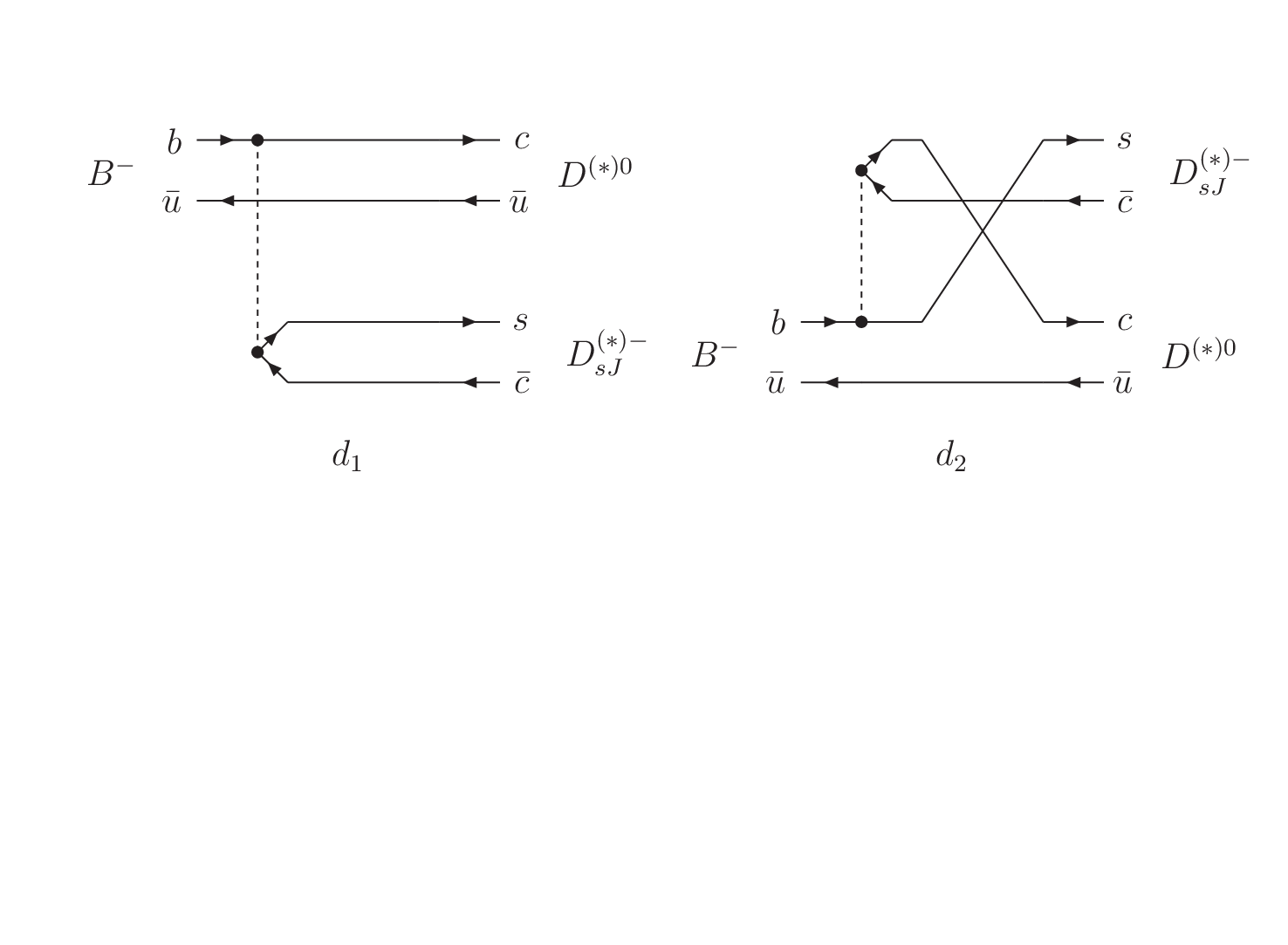,height=4.5cm,width=12cm}
\end{center}
\caption[Schematic representation of $B^{-}$ decay into $D^{0}D_{sJ}^{-}$]
{\label{fig:factBtoDDs} Schematic representation of $B^{-}$ decay into
$D^{(\ast)0}D_{sJ}^{-}$.}
\end{figure*}

We show in Fig.~\ref{fig:factBtoDDs} the schematic representation of the
$B^{-} \to D^{(\ast)0}D_{sJ}^{(\ast)-}$ decay given by the two local four-quark
operators, diagram $d_{1}$ for $Q_{1}^{cb}$ and diagram $d_{2}$ for
$Q_{2}^{cb}$. Factorization approximation is implicit in  diagram $d_{1}$, which
amounts to evaluate the hadron matrix element of the effective Hamiltonian as a
product of quark-current matrix elements. Fierz reordering of  diagram $d_{2}$
leads to the same contribution but for a colour factor. The full transition
amplitude can be evaluated with the $Q_1^{cb}$ part of the Hamiltonian but with 
an effective coupling given by
\begin{equation}
a_{1}(\mu') = C_1(\mu')+\frac{1}{N_{C}}\,C_{2}(\mu'),
\end{equation}
with $N_{C}=3$ the number of colors.

The decay width is given by
\begin{equation}
\begin{split}
\Gamma =& \frac{G_{F}^{2}}{16\pi m_{B}^{2}} \, |V_{cb}|^{2} \, |V_{cs}|^{2}
\, a_{1}^{2} \,
\frac{\lambda^{1/2}(m_{B}^{2},m_{D^{(\ast)}}^{2},m_{D_{sJ}^{(\ast)}}^{2})}{2m_{B
}} \\
&
\times {\cal H}_{\alpha\beta}(P_{B},P_{D^{(\ast)}}) \, \hat{\cal
H}^{{\alpha\beta}}(P_{D_{sJ}^{(\ast)}}),
\end{split}
\end{equation}
where $G_F=1.16637(1)\times10^{-5}\,\mbox{GeV}^{-2}$ is the Fermi decay
constant~\cite{PDG2010}, $\lambda(a,b,c)=(a+b-c)^2-4ab$, $V_{cb}$ and $V_{cs}$
are the corresponding Cabbibo\,-\,Kobayashi\,-\,Maskawa matrix elements, for
which we take $V_{bc}=0.0413$ and $V_{cs}=0.974$. $P_B$, $P_{D^{(\ast)}}$
and $P_{D_{sJ}^{(\ast)}}$ are the meson momenta.
${\cal H}_{\alpha\beta}(P_{B},P_{D^{(\ast)}})$ is the
hadron tensor for the $B^{-} \to D^{(\ast)0}$ transition and $\hat{\cal
H}^{{\alpha\beta}}(P_{D_{sJ}^{(\ast)}})$ is the hadron tensor for the ${\rm
vacuum} \to D_{sJ}^{(\ast)-}$ transition. We have 
\begin{equation}
\hat{\cal H}^{{\alpha\beta}}(P_{D_{sJ}^{(\ast)}})=P_{D_{sJ}^{(\ast)}}^{\alpha}
P_{D_{sJ}^{(\ast)}}^{\beta} f_{D_{sJ}^{(\ast)}}^{2},
\end{equation}
for a scalar or pseudoscalar meson, and
\begin{equation}
\hat{\cal H}^{{\alpha\beta}}(P_{D_{sJ}^{(\ast)}})= (P_{D_{sJ}^{(\ast)}}^{\alpha}
P_{D_{sJ}^{(\ast)}}^{\beta} -m_{D_{sJ}^{(\ast)}}^{2} g^{\alpha\,\beta})
f_{D_{sJ}^{(\ast)}}^{2},
\end{equation}
for a vector or an axial-vector meson. In previous equations,
$f_{D_{sJ}^{(\ast)}}$ is the $D_{sJ}^{(\ast)}$ meson decay constant.

As in Refs.~\cite{PhysRevD.74.074008,segovia2011semileptonic}, the contraction
of hadron tensors, ${\cal H}_{\alpha\beta}(P_{B_c},P_{D^{(\ast)}})\hat{\cal
H}^{{\alpha\beta}}(P_{D_{sJ}^{(\ast)}})$, can be easily written in terms of
helicity amplitudes for the $B^{-} \to D^{(\ast)}$
transition~\cite{PhysRevD.73.054024}, so that the decay width is given as
\begin{equation}
\begin{split}
\Gamma =& \frac{G_{F}^{2}}{16\pi m_{B}^{2}} |V_{cb}|^{2} |V_{cs}|^{2} a_{1}^{2}
\,
\frac{\lambda^{1/2}(m_{B}^{2},m_{D^{(\ast)}}^{2},m_{D_{sJ}^{(\ast)}}^{2})}{2m_{B
}} \times \\
& 
\times m_{D_{sJ}^{(\ast)}}^{2}f_{D_{sJ}^{(\ast)}}^{2}{\cal H}_{tt}^{B \to
D^{(\ast)}}(m_{D_{sJ}^{(\ast)}}^{2}),
\label{eq:gammaBDDs1}
\end{split}
\end{equation}
for $D_{sJ}^{(\ast)}$ a pseudoscalar or scalar meson, and
\begin{equation}
\begin{split}
\Gamma =& \frac{G_{F}^{2}}{16\pi m_{B}^{2}} |V_{cb}|^{2} |V_{cs}|^{2} a_{1}^{2}
\,
\frac{\lambda^{1/2}(m_{B}^{2},m_{D^{(\ast)}}^{2},m_{D_{sJ}^{(\ast)}}^{2})}{2m_
{B } } \times \\
&
\times m_{D_{sJ}^{(\ast)}}^{2} f_{D_{sJ}^{(\ast)}}^{2} \left[{\cal
H}_{+1+1}^{B\to D^{(\ast)}}(m_{D_{sJ}^{(\ast)}}^{2}) \right. \\
&
\left. +{\cal H}_{-1-1}^{B\to D^{(\ast)}}(m_{D_{sJ}^{(\ast)}}^{2})+{\cal
H}_{00}^{B\to D^{(\ast)}}(m_{D_{sJ}^{(\ast)}}^{2})\right],
\label{eq:gammaBDDs2}
\end{split}
\end{equation}
for $D_{sJ}^{(\ast)}$ a vector or axial-vector meson. The different ${\cal
H}_{rr}$ are defined in Ref.~\cite{PhysRevD.74.074008} and evaluated in this
case at $q^{2}=m_{D_{sJ}^{(\ast)}}^{2}$. The expressions of the decay
constants needed are given in the Appendix.

\section{RESULTS}
\label{sec:results}

In this section we present our results for the ratios in
Eqs.~(\ref{eq:DDsratio1}) and~(\ref{eq:DDsratio2}). We have calculated them
using the factorization approximation. As we have already mentioned, the
factorization approximation and heavy quark limit predicts ratios of order one
for $D_{s0}^{\ast}(2317)$ and $D_{s1}(2460)$ mesons, and negligible for
$D_{s1}(2536)$~\cite{LeYaouanc1996582,Datta2003164}.

>From Eqs.~(\ref{eq:gammaBDDs1}) and~(\ref{eq:gammaBDDs2}) we arrive at
\begin{widetext}
\begin{equation}
\begin{split}
R_{D0} =&
\frac{\lambda^{1/2}\left(m_{B}^{2},m_{D}^{2},m_{D_{s0}^{\ast}(2317)}^{2}
\right)m_ {D_{s0}^{\ast}(2317)}^{2}f_{D_{s0}^{\ast}(2317)}^{2}{\cal H}_{tt}^{B
\to D} \left(m_{D_{s0}^{\ast}(2317)}^{2}\right)}
{\lambda^{1/2}(m_{B}^{2},m_{D}^{2},m_{D_{s}}^{2})m_{D_{s}}^{2}f_{D_{s}}^{2}{\cal
H}_{tt}^{B \to D}(m_{D_{s}}^{2})}, \\[3ex]
R_{D1} =&
\frac{\lambda^{1/2}\left(m_{B}^{2},m_{D}^{2},m_{D_{s1}(2460)}^{2}\right) m_{D_
{s1}(2460)}^{2}f_{D_{s1}(2460)}^{2}}
{\lambda^{1/2}(m_{B}^{2},m_{D}^{2},m_{D_{s}^{\ast}}^{2})m_{D_{s}^{\ast}}^{2}f_{
D_{s}^{\ast}}^{2}} \times \\
&
\times \frac{\left[{\cal H}_{+1+1}^{B \to D}
\left(m_{D_{s1}(2460)}^{2}\right)+{\cal H}_{-1-1}^{B \to D}
\left(m_{D_{s1}(2460)}^{2}\right)+{\cal H}_{00}^{B \to
D}\left(m_{D_{s1}(2460)}^{2}\right)\right]} {\left[{\cal H}_{+1+1}^{B \to
D}(m_{D_{s}^{\ast}}^{2})+{\cal H}_{-1-1}^{B \to D}(m_{D_{s}^{\ast}}^{2})+{\cal
H}_{00}^{B \to D}(m_{D_{s}^{\ast}}^{2})\right]}, \\[3ex]
R_{D1'} =&
\frac{\lambda^{1/2}\left(m_{B}^{2},m_{D}^{2},m_{D_{s1}(2536)}^{2}\right) m_{D_
{s1}(2536)}^{2}f_{D_{s1}(2536)}^{2}}
{\lambda^{1/2}(m_{B}^{2},m_{D}^{2},m_{D_{s}^{\ast}}^{2})m_{D_{s}^{\ast}}^{2}f_{
D_{s}^{\ast}}^{2}} \times \\
&
\times \frac{\left[{\cal H}_{+1+1}^{B \to D}
\left(m_{D_{s1}(2536)}^{2}\right)+{\cal H}_{-1-1}^{B \to D}
\left(m_{D_{s1}(2536)}^{2}\right)+{\cal H}_{00}^{B \to
D}\left(m_{D_{s1}(2536)}^{2}\right)\right]} {\left[{\cal H}_{+1+1}^{B \to
D}(m_{D_{s}^{\ast}}^{2})+{\cal H}_{-1-1}^{B \to D}(m_{D_{s}^{\ast}}^{2})+{\cal
H}_{00}^{B \to D}(m_{D_{s}^{\ast}}^{2})\right]},
\end{split}
\end{equation}
\end{widetext}
and the same for $R_{D^{\ast}0}$, $R_{D^{\ast}1}$ and $R_{D^{\ast}1'}$ but
replacing the meson $D$ by the meson $D^{\ast}$.

Using experimental masses we obtain the ratios
\begin{equation}
\begin{split}
R_{D0} &= 0.9008 \times \left| \frac{f_{D_{s0}^{\ast}(2317)}}{f_{D_{s}}}
\right|^{2}, \\
R_{D^{\ast}0} &= 0.7166 \times \left| \frac{f_{D_{s0}^{\ast}(2317)}}{f_{D_{s}}}
\right|^{2}.
\label{eq:ratiosDs0}
\end{split}
\end{equation}
The double ratio $R_{D^{\ast}0}/R_{D0}$ does not depend on decay constants, and
in our model we obtain $R_{D^{\ast}0}/R_{D0}=0.7955$. The experimental value is
given by $R_{D^{\ast}0}/R_{D0}=1.50\pm0.75$. Our result is small compared to the
central experimental value but we are compatible within $1\sigma$. In the case
of the meson $D_{s1}(2460)$ we obtain
\begin{equation}
\begin{split}
R_{D1} &= 0.7040 \times \left| \frac{f_{D_{s1}(2460)}}{f_{D_{s}^{\ast}}}
\right|^{2}, \\
R_{D^{\ast}1} &= 1.0039 \times \left| \frac{f_{D_{s1}(2460)}}{f_{D_{s}^{\ast}}}
\right|^{2},
\label{eq:ratiosDs1}
\end{split}
\end{equation}
and for the double ratio $R_{D^{\ast}1}/R_{D1}$ we get $1.4260$, which agrees
well with the experimental result $R_{D^{\ast}1}/R_{D1}=1.32\pm0.43$. Finally,
for the meson $D_{s1}(2536)$ we obtain
\begin{equation}
\begin{split}
R_{D1'} &= 0.6370 \times \left| \frac{f_{D_{s1}(2536)}}{f_{D_{s}^{\ast}}}
\right|^{2}, \\
R_{D^{\ast}1'} &= 0.9923 \times \left| \frac{f_{D_{s1}(2536)}}{f_{D_{s}^{\ast}}}
\right|^{2},
\label{eq:ratiosDs1p}
\end{split} 
\end{equation}
and for the double ratio $R_{D^{\ast}1'}/R_{D1'}$, our value is $1.5578$ which
in this case is $2\sigma$ above the experimental one, $0.90\pm0.27$. 

The quality of the experimental results does not allow to be very conclusive as
to the goodness of factorization approximation. But one can conclude from
Eqs.~(\ref{eq:ratiosDs0}),~(\ref{eq:ratiosDs1}) and~(\ref{eq:ratiosDs1p}) that
one cannot ignore, as done when using the infinite heavy quark mass limit, phase
space and weak matrix element corrections.

\begin{table*}[!t]
\begin{center}
\begin{tabular}{llll} 
\hline
\hline
\tstrut
Approach & $f_{D}$ (MeV) & $f_{D_{s}}$ (MeV) & $f_{D_{s}}/f_{D}$ \\[0.2ex]
\hline
\tstrut
Ours & $297.019^{(\dag)}$ & $416.827^{(\dag)}$ & $1.40^{(\dag)}$ \\
     & $214.613^{(\ddag)}$ & $286.382^{(\ddag)}$ & $1.33^{(\ddag)}$ \\[1.5ex]
Experiment & $206.7\pm8.9$ & $257.5\pm6.1$ & $1.25\pm0.06$ \\
Lattice (HPQCD+UKQCD) & $208\pm4$ & $241\pm3$ & $1.162\pm0.009$ \\
Lattice (FNAL+MILC+HPQCD) & $217\pm10$ & $260\pm10$ & $1.20\pm0.02$ \\
PQL & $197\pm9$ & $244\pm8$ & $1.24\pm0.03$ \\
QL (QCDSF) & $206\pm6\pm3\pm22$ & $220\pm6\pm5\pm11$ & $1.07\pm0.02\pm0.02$ \\
QL (Taiwan) & $235\pm8\pm14$ & $266\pm10\pm18$ & $1.13\pm0.03\pm0.05$ \\
QL (UKQCD) & $210\pm10^{+17}_{-16}$ & $236\pm8^{+17}_{-14}$ &
$1.13\pm0.02^{+0.04}_{-0.02}$ \\
QL & $211\pm14^{+2}_{-12}$ & $231\pm 12^{+6}_{-1}$ & $1.10\pm0.02$\\
QCD Sum Rules & $177\pm21$ & $205\pm22$ & $1.16\pm0.01\pm0.03$ \\
QCD Sum Rules & $203\pm20$ & $235\pm 24$ & $1.15\pm 0.04$ \\
Field Correlators & $210\pm10$ & $260\pm10$ & $1.24\pm0.03$ \\
Light Front & $206$ & $268.3\pm19.1$ & $1.30\pm0.04$ \\
\hline
\hline
\tstrut
Approach & $f_{D^{\ast}}$ (MeV) & $f_{D_{s}^{\ast}}$ (MeV) &
$f_{D_{s}^{\ast}}/f_{D^{\ast}}$ \\[0.2ex]
\hline
\tstrut
Ours & $247.865^{(\dag)}$ & $329.441^{(\dag)}$ & $1.33^{(\dag)}$ \\[1.5ex]
RBS & $340\pm22$ & $375\pm24$ & $1.10\pm0.06$ \\
RQM & $315$ & $335$ & $1.06$ \\
QL (Italy) & $234$ & $254$ & $1.04\pm0.01^{+2}_{-4}$ \\
QL (UKQCD) & $245\pm20^{+0}_{-2}$ & $272\pm16^{+0}_{-20}$ & $1.11\pm0.03$ \\
BS & $237$ & $242$ & $1.02$ \\
RM & $262\pm10$ & $298\pm11$ & $1.14\pm0.09$ \\
\hline
\hline
\end{tabular}
\caption[Comparison of decay constants for pseudoscalar and vector charmed
mesons] {\label{tab:DCmodels} Theoretical predictions of decay constants for
pseudoscalar and vector charmed mesons. The data have been taken from
Ref.~\cite{PDG2010} for pseudoscalar mesons and from Ref.~\cite{GuoLi2006492}
for vector mesons. PQL$\equiv$Partially-Quenched Lattice calculation,
QL$\equiv$Quenched Lattice calculations, RBS$\equiv$Relativistic Bethe-Salpeter,
RQM$\equiv$Relativistic Quark Model, BS$\equiv$Bethe-Salpeter Method and
RM$\equiv$Relativistic Mock
meson model.}
\end{center}
\end{table*}

The decay constants of pseudoscalar and vector mesons in charmed and
charmed-strange sectors are given in Table~\ref{tab:DCmodels}. We compare our
results with the experimental data and those predicted by different approaches
and collected in Refs.~\cite{PDG2010,GuoLi2006492}. Our original values are
those with the symbol $(\dag)$. The decay constants of vector mesons agree with
other approaches. In the case of the pseudoscalar mesons, the decay constants
are simply too large. The reason for that is the following: Our CQM presents an
OGE potential which has a spin-spin contact hyperfine interaction that is
proportional to a Dirac delta function, conveniently regularized, at the origin.
The corresponding regularization parameter was fitted to determine the hyperfine
splittings between the $n^{1}S_{0}$ and $n^{3}S_{1}$ states in the different
flavor sectors, achieving a good agreement in all of them. While most of the
physical observables are insensitive to the regularization of this delta term,
those related with annihilation processes are affected because these processes
are driven by short range operators~\cite{Segovia:2011tb}. The effect is very
small in the $^{3}S_{1}$ channel as the delta term is repulsive in this case. It
is negligible for higher partial waves due to the shielding by the centrifugal
barrier. However, it is sizable in the $^{1}S_{0}$ channel for which the  delta
term is attractive. 

One should expect the wave functions of the $1^{1}S_{0}$ and $1^{3}S_{1}$
states to be very similar~\cite{MB2008455} except for the very short range. In
fact, they are equal if the Dirac delta term is ignored. The values with the
symbol $(\ddag)$ in Table~\ref{tab:DCmodels} are referred to the pseudoscalar
decay constants which have been calculated using the wave function of the
corresponding $^{3}S_{1}$ state. We recover the agreement with experiment and
also with the predictions of different theoretical approaches. The
$f_{D_{s}}/f_{D}$ and $f_{D_{s}^{\ast}}/f_{D^{\ast}}$ ratios are also shown in
the last column of Table~\ref{tab:DCmodels}. They are not very sensitive to the 
delta term and our values agree nicely with experiment and the values obtained
in other approaches.

\begin{table}[!t]
\begin{center}
\begin{tabular}{cccc}
\hline
\hline
\tstrut
Meson & & $f_{D}$ (MeV) & $\sqrt{M_{D}}f_{D}$ $({\rm GeV}^{3/2})$ \\
\hline
\tstrut
$D_{s0}^{\ast}(2317)$ & & $118.706$ & $0.181$ \\
$D_{s1}(2460)$ & & $165.097$ & $0.259$ \\
$D_{s1}(2536)$ & & $59.176$ & $0.094$ \\
\hline
\hline
\end{tabular}
\caption[Decay constants of some charmed-strange mesons]
{\label{tab:decayconstants} Decay constants calculated within the CQM including
1-loop QCD corrections to the OGE potential and non-$q\bar{q}$ structure in
channel $1^{+}$.}
\end{center}
\end{table}

Table~\ref{tab:decayconstants} summarizes the remaining decay constants needed
for the calculation we are interested in. There, we show the results from the
constituent quark model in which the 1-loop QCD corrections to the OGE potential
and the presence of non-$q\bar{q}$ degrees of freedom in $J^{P}=1^{+}$
charmed-strange meson sector are included. 

Refs.~\cite{Hwang2005116} and~\cite{PhysRevD.74.034020} calculate the lower
bounds of the decay constants of $D_{s0}^{\ast}(2317)$ and $D_{s1}(2460)$
analyzing experimental data related to $B\to DD_{sJ}^{(\ast)}$.
Ref.~\cite{Hwang2005116} provides the following lower limits:
\begin{equation}
\begin{split}
&
|a_{1}|\,f_{D_{s0}^{\ast}(2317)}=\begin{cases} 58-83\,{\rm MeV} &
\mbox{from }B^{-}\mbox{ decays} \\  63-86\,{\rm MeV} &
\mbox{from }\bar{B}^{0}\mbox{ decays} \end{cases} \\
&
|a_{1}|\,f_{D_{s1}(2460)}=\begin{cases} 188^{+40}_{-54}\,{\rm MeV} &
\mbox{from }B^{-}\mbox{ decays} \\  152^{+43}_{-62}\,{\rm MeV} &
\mbox{from }\bar{B}^{0}\mbox{ decays} \end{cases}
\end{split}
\end{equation}
and the authors of Ref.~\cite{PhysRevD.74.034020} get
\begin{equation}
\begin{split}
&
|a_{1}|\,f_{D_{s0}^{\ast}(2317)}=74\pm11 \\
&
|a_{1}|\,f_{D_{s1}(2460)}=166\pm20
\end{split}
\end{equation}
where the parameter $|a_{1}|\sim1$. Our results are compatible with these lower
limits.

Our results for the decay constants clearly deviate from the ones obtained in
the infinite heavy quark mass limit. In that limit one gets
$f_{D_{s0}^{\ast}(2317)} = f_{D_{s}}$, $f_{D_{s1}(2460)} = f_{D_{s}^{\ast}}$ and
$f_{D_{s1}(2536)} = 0$, results that lead to a strong disagreement with
experiment for the decay width ratios in Eqs.~(\ref{eq:DDsratio1})
and~(\ref{eq:DDsratio2}). That was already noticed in Ref.~\cite{Datta2003164},
where the authors, using the experimental ratios, estimated that
$f_{D_{s0}^{\ast}(2317)}\sim\frac{1}{3}f_{D_{s}}$ and
$f_{D_{s0}^{\ast}(2317)}\sim f_{D_{s1}(2460)}$ instead. We obtain
$f_{D_{s0}^{\ast}(2317)}/f_{D_{s}}=0.36$, $f_{D_{s0}^{\ast}(2317)} \sim
0.72 f_{D_{s1}(2460)}$ and $f_{D_{s1}(2536)}=59.176\,{\rm MeV}$, the latter
being small compared to the others but certainly different from zero.

\begin{table*}[!t]
\begin{center}
\begin{tabular}{lc|cccccccc}
\hline
\hline
\tstrut
& & \multicolumn{2}{c}{$X\equiv D_{s0}^{\ast}(2317)$} &
\multicolumn{3}{c}{$X\equiv D_{s1}(2460)$} &
\multicolumn{3}{c}{$X\equiv D_{s1}(2536)$} \\
& & The. & Exp. & & The. & Exp. & & The. & Exp. \\
\hline
\tstrut
${\cal B}(B \to DX)/{\cal B}(B \to DD_{s})$ & & $0.19^{(\ast)}$ &
$0.10\pm0.03$ & & - & - & & - & - \\[2ex]
${\cal B}(B \to D^{\ast}X)/{\cal B}(B \to D^{\ast}D_{s})$
& & $0.15^{(\ast)}$ & $0.15\pm0.06$ & & - & - & & - & - \\[2ex]
${\cal B}(B \to DX)/{\cal B}(B \to DD_{s}^{\ast})$ & & - & - &
& $0.177$ & $0.44\pm0.11$ & & $0.021$ & $0.049\pm0.010$ \\[2ex]
${\cal B}(B \to D^{\ast}X)/{\cal B}(B \to
D^{\ast}D_{s}^{\ast})$ & & - & - & & $0.252$ & $0.58\pm0.12$ & & $0.032$ &
$0.044\pm0.010$ \\[1ex]
\hline
\hline 
\end{tabular}
\caption[Ratios of branching fractions for nonleptonic decays 
$B\to D^{(\ast)}D_{sJ}^{(\ast)}$] {\label{tab:Ourratios} Ratios of branching fractions
for nonleptonic decays $B\to D^{(\ast)}D_{sJ}^{(\ast)}$.}
\end{center}
\end{table*}

Finally, we show in Table~\ref{tab:Ourratios} our results for the ratios written
in Eqs.~(\ref{eq:DDsratio1}) and~(\ref{eq:DDsratio2}). The symbol $(\ast)$
indicates that the ratios have been calculated using the experimental
pseudoscalar decay constant in Table~\ref{tab:DCmodels}. We get results close to
or within the experimental error bars for the $D_{s0}^{\ast}(2317)$ meson, which
to us is an indication that this meson could be a canonical $c\bar{s}$ state.
The incorporation of the non-$q\bar{q}$ degrees of freedom in the $J^{P}=1^{+}$
channel, enhances the $j_{q}=3/2$ component of the $D_{s1}(2536)$ meson and it
gives rise to ratios in better agreement with experiment. Note that this state
is still an almost pure $q\bar{q}$ state in our
description~\cite{PhysRevD.80.054017}.

The situation is more complicated for the $D_{s1}(2460)$ meson. The probability
distributions of its $^{1}P_{1}$ and $^{3}P_{1}$ components are corrected by the
inclusion of non-$q\bar{q}$ degrees of freedom, the latter making a $\sim25\%$
of the wave function. In our calculation, only the pure $q\bar{q}$ component 
of the $D_{s1}(2460)$ meson has been used to evaluate the $\Gamma(B\to
D^{(\ast)}D_{s1}(2460))$ decay width. The values we get for the corresponding
ratios in Eqs.~(\ref{eq:DDsratio1}) are lower than experiment.

\section{CONCLUSIONS}
\label{sec:conclusions}

We have performed a calculation of the ratios shown in Eqs.~(\ref{eq:DDsratio1})
and~(\ref{eq:DDsratio2}) working within the framework of the constituent quark
model and in the factorization approximation. These ratios have been
recently reported by the Belle Collaboration in Ref.~\cite{PhysRevD.83.051102}
and provide important information to check the structure of the
$D_{s0}^{\ast}(2317)$, $D_{s1}(2460)$ and $D_{s1}(2536)$ mesons.

The strong disagreement found between the heavy quark limit predictions and the
experimental data motivates the introduction of the finite $c$-quark mass
effects, which is done easily in the context of the constituent quark model.

In the heavy quark limit and in the factorization approximation, the ratios
should be of order one for the $D_{s0}^{\ast}(2317)$ and $D_{s1}(2460)$ mesons
and very small for $D_{s1}(2536)$. In contrast, the experimental data indicate
that the decay pattern of the $D_{s1}(2536)$ follows the expectations, but that
is not the case for the $D_{s0}^{\ast}(2317)$ and $D_{s1}(2460)$.

The mass of the $D_{s0}^{\ast}(2317)$ meson is lowered towards the experimental
value with the inclusion of the 1-loop corrections to the OGE potential. We also
obtain ratios compatible with the experimental data. Our results indicate that
this meson could be described as a canonical $c\bar s$ state.

We incorporate the non-$q\bar{q}$ degrees of freedom in the $J^{P}=1^{+}$
channel. The $D_{s1}(2536)$ meson remains almost a pure $q\bar{q}$ state and its
$j_{q}=3/2$ component is enhanced. This gives the correct ratios for the
$D_{s1}(2536)$ meson. Together with the properties calculated in
Refs.~\cite{PhysRevD.80.054017,segovia2011semileptonic}, this is to us evidence
of a good description of the $D_{s1}(2536)$ meson.

The $D_{s1}(2460)$ has a sizable non-$q\bar{q}$ component which contributes to
the decays under study. This contribution has not been calculated, as goes
beyond the scope of this work. We have computed the ratios considering only
the contribution coming from the $q\bar{q}$ structure of the $D_{s1}(2460)$
meson. The ratios are a factor 2 below the experimental ones.

\begin{acknowledgments}

This work has been partially funded by Ministerio de Ciencia y Tecnolog\'ia
under Contract Nos.  FPA2010-21750-C02-02 and FIS2011-28853-C02-02, by the
European Community-Research Infrastructure Integrating Activity 'Study of
Strongly Interacting Matter' (HadronPhysics2 Grant No. 227431) and by the
Spanish Ingenio-Consolider 2010 Program CPAN (CSD2007-00042). C. A. thanks a
Juan de la Cierva contract from the Spanish Ministerio de Educaci\'on y Ciencia.

\end{acknowledgments}

\begin{widetext}

\appendix

\section{DECAY CONSTANTS}
\label{appendix:decayconstants}

In our model, and due to the normalization of our nonrelativistic meson states,
the decay constants are given by~\cite{PhysRevD.71.113006}
\begin{equation}
\begin{split}
f_{M(0^{-})} &= -i \sqrt{\frac{2}{m_{M(0^{-})}}} \, \left\langle\right.\! 0 |
J_{A0}^{f_{2}f_{1}}(0) | M(0^{-}),\vec{0} \!\left.\right\rangle, \\
f_{M(0^{+})} &= +i \sqrt{\frac{2}{m_{M(0^{+})}}} \, \left\langle\right.\! 0 |
J_{V0}^{f_{2}f_{1}}(0) | M(0^{+}),\vec{0} \!\left.\right\rangle, \\
f_{M(1^{-})} &= - \sqrt{\frac{2}{m_{M(1^{-})}}} \, \left\langle\right.\! 0 |
J_{V3}^{f_{2}f_{1}}(0) | M(1^{-}),0\,\vec{0} \!\left.\right\rangle, \\
f_{M(1^{+})} &= + \sqrt{\frac{2}{m_{M(1^{+})}}} \, \left\langle\right.\! 0 |
J_{A3}^{f_{2}f_{1}}(0) | M(1^{+}),0\,\vec{0} \!\left.\right\rangle, \\
\end{split}
\end{equation}
for pseudoscalar, scalar, vector and axial-vector mesons, respectively.
$J_{V\mu}^{f_{2}f_{1}}$ and $J_{A\mu}^{f_{2}f_{1}}$ are the vector and
axial-vector charged weak current operators for a specific pair of quark flavors
$f_{1}$ and $f_{2}$.

The corresponding matrix elements are given by
\begin{equation*}
\begin{split}
\left\langle\right.\! 0 | J_{A0}^{f_{2}f_{1}}(0) | M(0^{-}),\vec{0}
\!\left.\right\rangle
=&
i\frac{\sqrt{3}}{\pi}\int d|\vec{p}\,|\,\,|\vec{p}\,|^{2} \hat
\phi^{(M(0^-))}_{f_1,f_2}(|\vec{p}\,|)\sqrt{\frac{(E_{f_{1}}(-\vec{p}\,)+m_{f_{1
}})(E_{f_{2}}(\vec{p}\,)+m_{f_{2}})}{4E_{f_1}(-\vec{p}\,)E_{f_2}(\vec{p}\,)}}
\\
&
\times \left[1-\frac{|\vec{p}\,|^{2}}{(E_{f_{1}}
(-\vec{p}\,)+m_{f_{1}})(E_{f_{2}}(\vec{ p}\,)+m_{f_{2}})}\right],
\end{split}
\end{equation*}

\begin{equation}
\begin{split}
\left\langle\right.\! 0 | J_{V0}^{f_{2}f_{1}}(0) | M(0^{+}),\vec{0}
\!\left.\right\rangle =&
i\frac{\sqrt{3}}{\pi}\int d|\vec{p}\,|\, |\vec{p}\,|^{3}
\hat{\phi}_{f_{1},f_{2}}^{(M(0^{+}))}(|\vec{p}\,|) \sqrt{\frac{(E_{f_{1}
}(-\vec{p}\,)+m_{f_{1}})(E_{f_{2}}(\vec{p}\,)+m_{f_{2}})}{4E_{f_1}(-\vec{p}\,)E_
{f_2}(\vec{p}\,)}} \\
&
\times \left[\frac{1}{E_{f_{2}}(\vec{p}\,)+m_{f_{2}}} -
\frac{1}{E_{f_{1}}(-\vec{p}\,)+m_{f_{1}}}\right], \\[2ex]
\left\langle\right.\! 0 | J_{V3}^{f_{2}f_{1}}(0) |
M(1^{-},L=0),0\,\vec{0} \!\left.\right\rangle = &
-\frac{\sqrt{3}}{\pi}\int
d|\vec{p}\,|\,|\vec{p}\,|^{2}\hat{\phi}_{f_{1},f_{2}}^{(M(1^{-},L=0))}
(|\vec{p}\,|)
\sqrt{\frac{(E_{f_{1}}(-\vec{p}\,)+m_{f_{1}})(E_{f_{2}}(\vec{p}\,)+m_{f_{
2}})}{4E_{f_1}(-\vec{p}\,)E_{f_2}(\vec{p}\,)}} \\
&
\times
\left(1+\frac{|\vec{p}\,|^{2}}{3(E_{f_{1}}(-\vec{p}\,)+m_{f_{1}})(E_{f_{2}
}(\vec {p}\,)+m_{f_{2}})}\right), \\[2ex]
\left\langle\right.\! 0 | J_{V3}^{f_{2}f_{1}}(0) |
M(1^{-},L=2,0)\,\vec{0} \!\left.\right\rangle = &
-\frac{2}{\pi}\sqrt{\frac{2}{3}}\int
d|\vec{p}\,| \, \hat{\phi}_{f_{1},f_{2}}^{(M(1^{-},L=2))}(|\vec{p}\,|)
\sqrt{\frac{(E_{f_{2}}(\vec{p}\,)+m_{f_{2}})(E_{f_{1}}(-\vec{p}\,)+m_{f_{1}})}{
4E_{f_1}(-\vec{p}\,)E_{f_2}(\vec{p}\,)}} \\
&
\times
\frac{|\vec{p}\,|^{4}}{(E_{f_{2}}(\vec{p}\,)+m_{f_{2}})(E_{f_{1}}(-\vec{p} \,)
+ m_{f_{1}})}, \\[2ex]
\left\langle\right.\! 0 | J_{A3}^{f_{2}f_{1}}(0) |
M(1^{+},S=0),0\,\vec{0} \!\left.\right\rangle = &
\frac{1}{\pi}\int
d|\vec{p}\,|\,|\vec{p}\,|^{3}\hat{\phi}_{f_{1},f_{2}}^{(M(1^{+},S=0))}
(|\vec{p}\,|)
\sqrt{\frac{(E_{f_{1}}(-\vec{p}\,)+m_{f_{1}})(E_{f_{2}}(\vec{p}\,)+m_{f_{
2}})}{4E_{f_1}(-\vec{p}\,)E_{f_2}(\vec{p}\,)}} \\ 
& 
\times 
\left[\frac{1}{E_{f_{1}}(-\vec{p}\,)+m_{f_{1}}}-\frac{1}{E_{f_{2}}(\vec{p}\,)+m_
{f_{2}}}\right], \\[2ex]
\left\langle\right.\! 0 | J_{A3}^{f_{2}f_{1}}(0) | M(1^{+},S=1),0\,\vec{0}
\!\left.\right\rangle =& -\frac{\sqrt{2}}{\pi}\int d|\vec{p}\,|\,|\vec{p}\,|^{3}
\hat{\phi}_{f_{1},f_{2}}^{(M(1^{+},S=1))}(|\vec{p}\,|)
\sqrt{\frac{(E_{f_{1}}(-\vec{p}\,)+m_{f_{1}})(E_{f_{2}}(\vec{p}\,)+m_{f_{2}})}{
4E_{f_1}(-\vec{p}\,)E_ {f_2}(\vec{p}\,)}} \\
&
\times
\left[\frac{1}{E_{f_{1}}(-\vec{p}\,)+m_{f_{1}}}+\frac{1}{E_{f_{2}}(\vec{p}\,)+m_
{f_{2}}}\right],
\label{eq:decaycte}
\end{split}
\end{equation}
where $E_{f_1}$ and $E_{f_2}$ are the relativistic energy of the quarks. For
$0^{-}$ and $0^{+}$ we have only one possible contribution. In the case of the
$J^{P}=1^{-}$ meson we have two contributions coming from the two possible
values of the relative angular momentum. For $J^{P}=1^{+}$ states and since
$C$-parity is not well defined in charmed and charmed-strange mesons the wave
function is a mixture of $^{1}P_{1}$ and $^{3}P_{1}$ partial waves and thus
there are also two contributions.

\subsection{Decay constants in the heavy quark limit}

For the low-lying positive parity excitations, any quark model predicts four
states that in the $^{2S+1}L_{J}$ basis correspond to $^1P_1$, $^3P_0$, $^3P_1$
and $^3P_2$. As charge conjugation is not well defined in the heavy-light
sector, $^1P_1$ and $^3P_1$ states can mix under the interaction.

In the infinite heavy quark mass limit, heavy quark symmetry (HQS) predicts two
degenerate $P$-wave meson doublets, labeled by $j_{q}=1/2$ with
$J^{P}=0^{+},1^{+}$ $(|1/2,0^{+}\rangle,|1/2,1^{+}\rangle)$ and $j_{q}=3/2$ with
$J^{P}=1^+,2^+$ $(|3/2,1^+\rangle,|3/2,2^+\rangle)$. In this limit, the meson
properties are governed by the dynamics of the light quark, which is
characterized by its total angular momentum $j_{q}=s_{q}+L$, where $s_{q}$ is
the light quark spin and $L$ the orbital angular momentum. The total angular
momentum of the meson $J$ is obtained coupling $j_{q}$ to the heavy quark spin,
$s_{Q}$. 

A change of basis allows to express the above states in terms of the
$^{2S+1}L_{J}$ basis, by recoupling angular momenta, as
\begin{equation}
\begin{split}
&
|1/2,0^{+}\!\!\left.\right\rangle=+|^{3}P_{0}\!\!\left.\right\rangle, \\
&
\begin{split}
|1/2,1^{+}\!\!\left.\right\rangle &=
+\sqrt{\frac{1}{3}}|^{1}P_{1}\!\!\left.\right\rangle
+\sqrt{\frac{2}{3}}|^{3}P_{1}\!\!\left.\right\rangle,
\end{split} \\
&
\begin{split}
|3/2,1^{+} \!\!\left.\right\rangle &=
- \sqrt{\frac{2}{3}}|^{1}P_{1}\!\!\left.\right\rangle
+ \sqrt{\frac{1}{3}}|^{3}P_{1}\!\!\left.\right\rangle,
\end{split} \\
&
|3/2,2^{+} \!\!\left.\right\rangle=+|^{3}P_{2}\!\!\left.\right\rangle,
\end{split}
\label{eq:mixing}
\end{equation}
where in the $^{2S+1}L_J$ wave functions we couple heavy and light quark spins,
in this order, to total spin $S$.

In the actual calculation the ideal mixing in Eq.~(\ref{eq:mixing}) between
$^1P_1$ and $^3P_1$ states changes due to finite charm quark mass effects. Our
CQM model  predicts the mixed states shown in Table~\ref{tab:mixedstates}, which
are very similar to the HQS states. This is expected since the $c$\,-\,quark is
much heavier ($m_{c}=1763\,{\rm MeV}$) than the light ($m_{n}=313\,{\rm MeV}$)
or strange ($m_{s}=555\,{\rm MeV}$) quarks. Note that we also have mixing, even
if small, between the $^{3}P_{2}$ and $^{3}F_{2}$ partial waves in $2^{+}$
mesons. This is due to the OGE tensor term.

In Ref.~\cite{PhysRevD.80.054017} we have studied the $J^{P}=1^{+}$
charmed-strange mesons, finding that the $J^{P}=1^{+}$ $D_{s1}(2460)$ has an
important non-$q\bar q$ contribution whereas the $J^{P}=1^{+}$ $D_{s1}(2536)$ is
almost a pure $q\bar{q}$ state. The presence of non-$q\bar q$ degrees of freedom
in the $J^{P}=1^{+}$ charmed-strange meson sector enhances the $j_q=3/2$
component of the $D_{s1}(2536)$. This wave function explains most of the
experimental data, as shown in
Refs.~\cite{PhysRevD.80.054017,segovia2011semileptonic}. For this sector only
the $q\bar q$ probabilities are given in Table~\ref{tab:mixedstates}.

\begin{table}[t!]
\begin{center}
\begin{tabular}{ccccc}
\hline
\hline
\tstrut
& $D_{0}^{\ast}(2400)$ & $D_{1}(2420)$ & $D_{1}(2430)$ & $D_{2}^{\ast}(2460)$ \\
\hline
\tstrut
$^{3}P_{0}$ & $+,\,1.0000$ & - & - & - \\
$^{1}P_{1}$ & - & $-,\,0.5903$ & $-,\,0.4097$ & - \\
$^{3}P_{1}$ & - & $+,\,0.4097$ & $-,\,0.5903$ & - \\
$^{3}P_{2}$ & - & - & - & $+,\,0.99993$ \\[2ex]
$1/2,0^+$ & $+,\,1.0000$ & - & - & - \\
$1/2,1^+$ & - & $+,\,0.0063$ & $-,\,0.9937$ & - \\
$3/2,1^+$ & - & $+,\,0.9937$ & $+,\,0.0063$ & - \\
$3/2,2^+$ & - & - & - & $+,\,0.99993$ \\
\hline
\hline
\tstrut
& $D_{s0}^{\ast}(2317)$ & $D_{s1}(2536)$ & $D_{s1}(2460)$ &
$D_{s2}^{\ast}(2573)$ \\
\hline
\tstrut
& $D_{s0}^{\ast}$ & $D_{s1}$ & $D'_{s1}$ & $D_{s2}^{\ast}$ \\[2ex]
$^{3}P_{0}$ & $+,\,1.0000$ & - & - & - \\
$^{1}P_{1}$ & - & $-,\,0.7210$ & $-,\,0.1880$ & - \\
$^{3}P_{1}$ & - & $+,\,0.2770$ & $-,\,0.5570$ & - \\
$^{3}P_{2}$ & - & - & - & $+,\,0.99991$ \\[2ex]
$1/2,0^+$ & $+,\,1.0000$ & - & - & - \\
$1/2,1^+$ & - & $-,\,0.0038$ & $-,\,0.7390$ & - \\
$3/2,1^+$ & - & $+,\,0.9942$ & $-,\,0.0060$ & - \\
$3/2,2^+$ & - & - & - & $+,\,0.99991$ \\
\hline
\hline
\end{tabular}
\caption{\label{tab:mixedstates} Probability distributions  and their
relative phases for the four states predicted by CQM in the two basis
described in the text. In the $1^+$ strange sector the effects of
non-$q\bar q$ components are included, see text for details.}
\end{center}
\end{table}

Here, we want to recover the results of the decay constants predicted by HQS.
>From Eqs.~(\ref{eq:decaycte}), we only need to do the limit $m_{f_{1}}\to\infty$
\begin{itemize}
\item Decay constants of $D_{s}$ and $D_{s}^{\ast}$ in the heavy quark limit:
\begin{equation}
\begin{split}
f_{D_{s}} \xrightarrow{m_{f_{1}}\to\infty} &
\frac{\sqrt{3}}{\pi}\sqrt{\frac{1}{m_{D_{s}}}} \int
d|\vec{p}\,|\,\,|\vec{p}\,|^{2}
\hat\phi^{(D_{s},^{1}S_{0})}_{f_1,f_2}(|\vec{p}\, |)
\sqrt{1+\frac{m_{f_{2}}}{E_{ f_{2} }(\vec{p}\,)}}\,\,, \\
f_{D_{s}^{\ast}} \xrightarrow{m_{f_{1}}\to\infty} & 
\frac{\sqrt{3}}{\pi}\sqrt{\frac{1}{m_{D_{s}^{\ast}}}} \int
d|\vec{p}\,|\,|\vec{p}\,|^{2}\hat{\phi}_{f_{1},f_{2}}^{(D_{s}^{\ast},^{3}S_{1})}
(|\vec{p}\,|) \sqrt{1+\frac{m_{f_{2}}}{E_{ f_{2} }(\vec{p}\,)}}\,\,.
\end{split}
\end{equation}
The $D_{s}$ and $D_{s}^{\ast}$ mesons belong to the
doublet $(|1/2,0^{-}\rangle,|1/2,1^{-}\rangle)$ predicted by HQS. In such a
limit the mesons are degenerate and the radial wave functions of the $^{1}S_{0}$
and $^{3}S_{1}$ components are the same. Therefore, we obtain
$f_{D_{s}}=f_{D_{s}^{\ast}}$ in that limit.
\item Decay constants of $D_{s0}^{\ast}(2317)$ and $D_{s1}(2460)$ in the heavy
quark limit:
\begin{equation}
\begin{split}
f_{D_{s0}^{\ast}(2317)} \xrightarrow{m_{f_{1}}\to\infty} &
-\frac{\sqrt{3}}{\pi} \sqrt{\frac{1}{m_{D_{s0}^{\ast}(2317)}}}
\int d|\vec{p}\,|\, |\vec{p}\,|^{3}
\hat{\phi}_{f_{1},f_{2}}^{(D_{s0}^{\ast}(2317),^{3}P_{0})}(|\vec{p}\,|)
\sqrt{1+\frac{m_{f_{2}}}{E_{f_{2}}(\vec{p}\,)}}\,\frac{1}{
E_{f_{2}} (\vec{p}\,)+m_{f_{2}}}, \\[3ex]
f_{D_{s1}(2460)} \xrightarrow{m_{f_{1}}\to\infty} &
+\sqrt{\frac{2}{m_{D_{s1}(2460)}}} \times \\
&
\times \left\lbrace -\sqrt{\frac{1}{6}}
\frac{1}{\pi}\int
d|\vec{p}\,|\,|\vec{p}\,|^{3}
\hat{\phi}_{f_{1},f_{2}}^{(D_{s1}(2460),^{1}P_{1})}
(|\vec{p}\,|) \sqrt{1+\frac{m_{f_{2}}}{E_{f_{2}}(\vec{p}\,)}}\, 
\frac{1}{E_{f_{2}}(\vec{p}\,)+m_{f_{2}}} \right. \\
&
\quad \quad \!\! \left. -\sqrt{\frac{2}{3}}\frac{1}{\pi}\int
d|\vec{p}\,|\,|\vec{p}\,|^{3}
\hat{\phi}_{f_{1},f_{2}}^{(D_{s1}(2460),^{3}P_{1})}(|\vec{p}\,|)
\sqrt{1+\frac{m_{f_{2}}}{E_{f_{2}}(\vec{p}\,)}}\,
\frac{1}{E_{f_ {2}}(\vec{p}\,)+m_{f_{2}}} \right\rbrace.
\end{split}
\end{equation}
The $D_{s0}^{\ast}(2317)$ and $D_{s1}(2460)$ mesons belong to the doublet
$(|1/2,0^{+}\rangle,|1/2,1^{+}\rangle)$ predicted by HQS. Again, in such a
limit, the mesons are degenerate and the radial wave functions of the
$^{3}P_{0}$, $^{1}P_{1}$ and $^{3}P_{1}$ components would be equal. Therefore,
we would obtain $f_{D_{s0}^{\ast}(2317)}=f_{D_{s1}(2460)}$ in that limit.

\item Decay constant of $D_{s1}(2536)$ in the HQS:
\begin{equation}
\begin{split}
f_{M(1^{+},D_{s1}(2536))} \xrightarrow{m_{f_{1}}\to\infty} &
+\sqrt{\frac{2}{m_{D_{s1}(2536)}}} \times \\ 
&
\times \left\lbrace +\sqrt{\frac{1}{3}} \frac{1}{\pi}\int
d|\vec{p}\,|\,|\vec{p}\,|^{3}
\hat{\phi}_{f_{1},f_{2}}^{(D_{s1}(2536),^{1}P_{1})}
(|\vec{p}\,|) \sqrt{1+\frac{m_{f_{2}}}{E_{f_{2}}(\vec{p}\,)}}\,
\frac{1}{E_{f_{2}}(\vec{p}\,)+m_{f_{2}}} \right. \\
&
\quad \quad \!\! \left. -\sqrt{\frac{1}{3}}\frac{1}{\pi}\int
d|\vec{p}\,|\,|\vec{p}\,|^{3}
\hat{\phi}_{f_{1},f_{2}}^{(D_{s1}(2536),^{3}P_{1})}(|\vec{p}\,|)
\sqrt{1+\frac{m_{f_{2}}}{E_{f_{2}}(\vec{p}\,)}}\,
\frac{1}{E_{f_ {2}}(\vec{p}\,)+m_{f_{2}}} \right\rbrace.
\end{split}
\end{equation}
Therefore, considering that the radial wave functions of the $^{1}P_{1}$ and
$^{3}P_{1}$ components are the same, $f_{D_{s1}(2536)}$ is equal zero in the
heavy quark limit.
\end{itemize}

\end{widetext}


\bibliographystyle{apsrev}
\bibliography{weak_paper2}

\end{document}